\documentclass{PoS}

\usepackage{amsmath,amssymb}
\usepackage{dsfont}
\usepackage{xspace}
\usepackage{tabularx}
\usepackage{subfigure}
\usepackage[sort&compress,numbers,merge]{natbib}
\usepackage{natbib,natbibspacing}
\usepackage[vlined,ruled,linesnumbered]{algorithm2e}
\usepackage{xfrac}
\newcommand{\SetAlgorithmStyle}{
  \SetKwInput{Input}{input}
  \SetKwInput{Output}{output}
  \SetKwComment{tcp}{\{}{\}}
  \SetKwComment{tcpS}{\{}{\phantom{\}}}
  \SetKwComment{tcpM}{}{\phantom{\}}}
  \SetKwComment{tcpE}{}{\}}
  \SetKwFor{ForAll}{for all}{do}{end}
  \SetKwFor{For}{for}{}{end}
  \SetKwFor{While}{while}{do}{end}
  \SetArgSty{}
  \DontPrintSemicolon
}\newcommand{\tr}{\operatorname{tr}}

\newcommand{\vect}[1]{\mathbf{#1}}
\newcommand{\proj}[1]{\mathcal{#1}}
\newcommand{\norm}[1]{\left\lVert#1\right\rVert}

\title{(Approximate) Low-Mode Averaging with a new Multigrid Eigensolver}

\ShortTitle{LMA with a new MG Eigensolver}

\author{Gunnar~Bali\textsuperscript{a}, Sara~Collins\textsuperscript{a}, Andreas~Frommer\textsuperscript{b}, Karsten~Kahl\textsuperscript{b}, Issaku~Kanamori\textsuperscript{c}, Benjamin~M\"uller\textsuperscript{d}, Matthias~Rottmann\textsuperscript{b,*}, Jakob~Simeth\textsuperscript{a,*} \phantom{\speaker{M.~Rottmann, J.~Simeth}}\\
  \textsuperscript{a}Institut f\"ur Theoretische Physik, Universit\"at Regensburg\\
  \textsuperscript{b}Fachbereich Mathematik und Naturwissenschaften, Bergische Universit\"at Wuppertal\\
  \textsuperscript{c}Department of Physics, Hiroshima University\\
  \textsuperscript{d}Institut f\"ur Mathematik, Johannes Gutenberg Universit\"at Mainz\\
  \email{rottmann@math.uni-wuppertal.de}, \email{jakob.simeth@physik.uni-regensburg.de}}

\abstract{We present a multigrid based eigensolver for computing low-modes of the Hermitian Wilson Dirac operator. For 
the non-Hermitian case multigrid methods have already replaced conventional Krylov subspace solvers in many lattice QCD 
computations. Since the $\gamma_5$-preserving aggregation based interpolation used in our multigrid method is valid for 
both, the Hermitian and the non-Hermitian case, inversions of very ill-conditioned shifted systems with the Hermitian 
operator become feasible. This enables the use of multigrid within shift-and-invert type eigensolvers. We show numerical 
results from our MPI-C implementation of a Rayleigh quotient iteration with multigrid. For state-of-the-art lattice 
sizes and moderate numbers of desired low-modes we achieve speed-ups of an order of magnitude and more over PARPACK. We 
show results and develop strategies how to make use of our eigensolver for calculating disconnected contributions to 
hadronic quantities that are noisy and still computationally challenging. Here, we explore the possible benefits, using 
our eigensolver for low-mode averaging and related methods with high and low accuracy eigenvectors. We develop a 
low-mode averaging type method using only a few of the smallest eigenvectors with low accuracy. This allows us to avoid 
expensive exact eigensolves, still benefitting from reduced statistical errors.
}

\FullConference{The 33rd International Symposium on Lattice Field Theory\\
		14 -18 July 2015\\
		Kobe International Conference Center, Kobe, Japan}

\begin{document}
\linespread{0.99}
\section{Introduction and Motivation}
There are many applications of eigensolvers in Lattice QCD: E.g.,~many physical properties are encoded in the spectrum 
of the Dirac operator, \(D\), and the lowest eigenvalues and eigenvectors of \(D^\dagger D\) can be used in low-mode 
averaging to reduce the noise of stochastically estimated quantities like disconnected fermion loops.
However, the calculation of the lowest eigenmodes of the Dirac operator can be costly and scales with \(V\,N_{eig}^2\), 
where \(V\) is the lattice four-volume and \(N_{eig}\) is the number of the lowest eigenmodes and often \(N_{eig}\propto 
V\).
There are two possible ways to alleviate this problem and make eigenmodes affordable for the use in low-mode averaging and other applications \cite{Neff:2001zr,DeGrand:2004qw,Giusti:2004yp,Bali:2005fu,Foley:2005ac}. One of them is the development of more efficient solvers, the other is to relax the precision of the eigenmodes. In this work we pursue both paths.

Several adaptive algebraic multigrid methods have been proposed in recent 
years as linear system solvers for the non-Hermitian formulation; 
cf.~\cite{MGClark2010_1,MGClark2007,MGClark2010_2,Frommer:2013fsa,Frommer:2013kla}. In particular, we proposed an 
adaptive aggregation based domain decomposition multigrid (``DD-$\alpha$AMG'') method to solve linear systems with the 
non-Hermitian Wilson Dirac operator $D$ and observed large speed-ups over conventional Krylov subspace methods. In what 
follows we present a modification of this method that allows us to also solve systems with the Hermitian version of the 
Dirac operator $Q = \gamma_5 D$. In order to use this linear systems solver to calculate eigenmodes of $Q$ we employ a 
standard Rayleigh quotient iteration~(see~\cite{saad2011numerical}), where shifted systems $Q-\sigma$ need to 
be inverted. In this process, the current eigenvector approximations are built into the interpolation in each iteration, 
which allows us to view the eigensolver also as a setup procedure for the multigrid method itself, and thus enables 
the calculation of eigenvectors corresponding to small eigenvalues to any desired accuracy. We compare a 
\texttt{MPI}-\texttt{C} implementation of our eigensolver with PARPACK and show that speed-ups of roughly an order of 
magnitude can be achieved.

Subsequently, we use this method to obtain both high and low accuracy eigenmodes of the Hermitian Dirac operator and use these in low-mode averaging which we apply to the pion- and eta-correlators. By constructing an improved estimate for approximate low-mode contributions we are able to benefit even more from the faster calculation when relaxing the target residual. The introduction of a cutoff enables us to use the test vectors from the standard DD-$\alpha$AMG setup~\cite{Frommer:2013fsa} without further iteration on the eigenvectors. By combining these two approaches, we obtain final statistical errors which are of roughly the same magnitude as those obtained when using exact eigenmodes but at a much smaller total cost.

The structure of this work is as follows: In Sec.~\ref{sec:algo} we describe our multigrid eigensolver algorithm for \(Q\) using Rayleigh quotient iteration and subsequently compare the performance to PARPACK in Sec.~\ref{sec:scaling}.
In Sec.~\ref{sec:physicsapp} we apply this method to low-mode averaging for the eta- and pion-correlator in the 
two-flavour system. After a short introduction to the main techniques, namely low-mode averaging and stochastic 
estimation, in Sec.~\ref{sec:lma} we present improvement techniques for the inexact eigenmodes (Sec.~\ref{sec:ainv}), and 
in Sec.~\ref{sec:cutoff} we devise a criterion to restrict the set of eigenmodes so that we can use the test vectors of 
the multigrid setup directly. Finally, we compare errors and the achieved speed-ups in Sec.~\ref{sec:results} before we 
conclude in Sec.~\ref{sec:conclusions}.
\section{Algebraic Multigrid in Rayleigh Quotient Iteration}
\label{sec:algo}
Let $D$ be the non-Hermitian (Clover-improved) Wilson Dirac operator and \(Q := \gamma_5 D\)
its Hermitian version ($\gamma_5 = - \gamma_1\gamma_2\gamma_3\gamma_4$) and assume that we choose a 
representation of the $\gamma_{i}$ such that $\gamma_{5} = \left(\begin{matrix}\mathds{1}&\\&-\mathds{1}\end{matrix}\right)$. An 
eigenvector $\mid v \rangle \neq 0$ of $Q$ with corresponding eigenvalue $\lambda$ satisfies
\begin{equation}
Q \mid v \rangle = \lambda \mid v \rangle \, .
\end{equation}  
If $\lambda$ is small in modulus, we call $ \mid v \rangle $ a \emph{small eigenvector} or \emph{low-mode}. In~\cite{Frommer:2013fsa,Frommer:2013kla} we proposed an adaptive algebraic domain decomposition method termed ``DD-$\alpha$AMG'' for solving linear systems
\begin{equation}
  D \mid \psi \rangle = \mid \eta \rangle \, .
\end{equation}
The error propagator for the two-level version of our method is -- as for many other two-level approaches -- of the 
generic form 
\begin{equation}
  E_{2g} = (\mathds{1}-MD)^{\nu}(\mathds{1}-P D_{c}^{-1} R D)(\mathds{1}-MD)^{\mu},
\end{equation}
where $M$ denotes the smoother which is given by the Schwarz alternating procedure (SAP), $\mu$ and $\nu$ number 
the pre- and post-smoothing iterations, respectively. $P$ denotes the adaptively constructed aggregation based 
interpolation / prolongation~\cite{Frommer:2013fsa,Frommer:2013kla,MGClark2010_2,MGClark2010_1,MGClark2007} and $R$ the 
corresponding restriction. In order to preserve the $\gamma_{5}$-hermiticity of $D$ in the coarse grid system 
$D_{c} := RDP$, as in the other approaches just cited, we choose $P = \left(\begin{matrix}P_{1}& 0\\ 0 &P_{2}\end{matrix}\right)$ in accordance to the ordering of 
spins in $\gamma_{5}$, so that $P$ fulfils
\begin{equation}
	\gamma_5 P = P \gamma_5^{c},
\end{equation} where $\gamma_{5}^{c}$ is the coarse grid analogue of $\gamma_{5}$.
Thus, further choosing $R = P^\dagger$ we obtain a $\gamma_5^c$ symmetric coarse operator that fulfils
\begin{equation}
	\gamma_5^{c} D_{c} = P^\dagger \gamma_5 D P = P^\dagger D^\dagger P \gamma_5^c = D_c^\dagger \gamma_5^c = ( \gamma_5^c D_c 
)^\dagger  \, .
\end{equation}

This algebraic multigrid approach for $D$ can then be easily transferred to one for $Q$. In fact it has been shown 
already in~\cite{MGClark2010_2} that using this construction for $P$ and $R$ the coarse grid corrections for $D$ and 
$Q$ are identical if one chooses $Q_{c} := P^\dagger Q P$, i.e., 
\begin{equation}
  \mathds{1}-P Q_{c}^{-1} P^\dagger Q = \mathds{1}-P D_c^{-1} \gamma_5^c P^\dagger \gamma_5 D = \mathds{1}-P D_c^{-1} P^\dagger D \, .
\end{equation}
The remaining part is to find a smoother for $Q$ and define a way to solve systems with the coarse grid system 
$Q_c$. Numerical experiments show that SAP is not suitable as a smoother for $Q$.
Since full GMRES is known to converge for any linear system, restarted GMRES in 
practice may work as a smoother for $Q$ for suitably chosen restart lengths. At the same time, 
GMRES is one of the most 
numerically stable Krylov subspace methods for indefinite systems and is thus to be expected to work well as a solver for $Q_c$. 
Therefore, we supply (restarted) GMRES 
as smoother and coarse grid solver which makes our algebraic multigrid solver for $Q$ similar in spirit to the 
one proposed 
in~\cite{MGClark2007,MGClark2010_1,MGClark2010_2} for $D$.
Due to the fact that the multigrid method is adaptively constructed to efficiently treat the low modes of $Q$ on the 
coarse grid, it should also do so for $Q-\sigma$ as long as $\sigma$ (in modulus) is sufficiently small. 
This then allows to construct shift-invert eigensolvers where algebraic multigrid accelerates 
the eigensolver. The simpler shift-invert approaches supply shifts $\sigma$ close to an eigenvalue $\lambda$ such 
that $Q-\sigma$ becomes very ill-conditioned. It has been shown 
in~\cite{Frommer:2013fsa,Frommer:2013kla,MGClark2010_2,MGClark2010_1,MGClark2007} that algebraic multigrid methods for 
$D$ are less sensitive to 
the condition number than Krylov subspace methods. Since this also holds for algebraic multigrid for $Q$, the coarse 
grid corrections being equivalent, it appears attractive to use algebraic multigrid for $Q$ in an eigensolver 
setting.

\subsection{Description of the Algorithm}

A standard shift-invert eigensolver approach is the Rayleigh 
quotient iteration (RQI, see, e.g.,~\cite{saad2011numerical}). We now describe RQI that uses 
our algebraic multigrid method for the inversion 
of the shifted systems in detail. We initially choose a set of $N_{eig}$ orthonormalised random vectors 
$ \mid v_1 \rangle ,\ldots, \mid v_{N_{eig}} \rangle $ and corresponding eigenvalue guesses $\lambda_1=\ldots=\lambda_N=0$. Using the multigrid method each 
$ \mid v_i \rangle $ is updated via one shifted inversion $ \mid v_i \rangle \leftarrow (Q-\lambda_i)^{-1} \mid v_i \rangle $. Subsequently, the 
vectors $ \mid v_1 \rangle ,\ldots, \mid v_{N_{eig}} \rangle $ are re-orthonormalised and the eigenvalue guesses $\lambda_i$ are updated as $\lambda_i = \langle v_i \mid Q \mid v_i \rangle $. 
This process is iterated until the norm of the eigenvector residual $\norm{Q \mid v_i \rangle - \lambda \mid v_i 
\rangle}$ is smaller than a given tolerance $\varepsilon$. As opposed to standard RQI, the solver that is used to 
solve the shifted linear systems is updated in every iteration by re-building the interpolation operator $P$ from the 
recent eigenvector estimates $ \mid v_1 \rangle ,\ldots, \mid v_{N_{eig}} \rangle $.

\begin{algorithm}[ptb]
  \label{alg:rqi_amg}
  \caption{Rayleigh Quotient Iteration + algebraic multigrid}\label{alg:evalDirac}
  \SetAlgorithmStyle
    \Input{number of eigenvectors $N_{eig}$, desired accuracy $\varepsilon$}
    \Output{eigenvectors $ \mid v_1 \rangle,\ldots, \mid v_{N_{eig}} \rangle $}    
    let $ \mid v_1 \rangle,\ldots, \mid v_{N_{eig}} \rangle $ be orthonormalised random vectors and $\lambda_i=0, 
\;\varepsilon_i=1 \; \forall i=1,\ldots,N_{eig}$\;
    build $P$ from $ \mid v_1 \rangle,\ldots, \mid v_{N_{eig}} \rangle $\;
    \While{$ \exists \varepsilon_i \, : \, \varepsilon_i > \varepsilon $}{
      \ForAll{$i=1,\ldots,N_{eig}$ with $\varepsilon_i > \varepsilon$}{
        $\sigma \leftarrow \lambda_i \cdot \mathrm{max}(1-\varepsilon_i,0)$\; \label{line:eigvalguess}
        $\mid v_i \rangle \leftarrow ( Q-\sigma )^{-1} \mid v_i \rangle$\; 
        $\mid v_i \rangle \leftarrow \mid v_i \rangle - \sum_{j=1}^{i-1} ( \langle v_j \mid v_i \rangle ) \mid v_j \rangle $\;
        $\mid v_i \rangle \leftarrow \mid v_i \rangle / \norm{ \mid v_i \rangle } $\;
        update $v_i$ in interpolation $P$\;
        $\lambda_i = \langle v_i \mid Q \mid v_i \rangle$\;
        $\varepsilon_i = \norm{Q \mid v_i \rangle - \lambda_i \mid v_i \rangle}$\; 
      }
    }
\end{algorithm}

The described procedure is summarised in Alg.~\ref{alg:rqi_amg}. In practice we do not start with entirely random 
vectors $\mid v_1 \rangle,\ldots, \mid v_{N_{eig}} \rangle$ but with the test vectors generated in the setup phase which constructs the 
interpolation $P$ used in DD-$\alpha$AMG. This setup procedure, described in~\cite{Frommer:2013fsa}, applies a small 
number of smoother iterations to a set of random vectors to form  a first interpolation $P$. Then, one iteration of 
algebraic multigrid is applied to each test vector while keeping them orthonormal. This procedure is repeated a few 
times until one is satisfied with the convergence of the algebraic multigrid solver.

In practice we observed that sometimes the $N_{eig}$ computed eigenpairs $(\lambda_i,\mid v_i \rangle)$ are not the $N_{eig}$ 
smallest pairs. Usually, the eigenpairs with $\lambda$ closest to 0 are always met, but some of the remaining smallest eigenvalues
might be missed. This can happen if the starting guess has only very little overlap with the direction of the desired eigenvector or if the estimate for the current eigenvalue is too large. In order to reduce the frequency of such events, we introduce an accuracy dependent damping mechanism in 
line~\ref{line:eigvalguess} that restricts the magnitude of the 
shifts used in the shifted inversion. 

\subsection{Scaling and Comparison with Other Algorithms}
\label{sec:scaling}
In this section we give preliminary results for our Rayleigh quotient iteration with multigrid algorithm (RQI+AMG, Algorithm~\ref{alg:rqi_amg}) that 
we implemented within our existing DD-$\alpha$AMG framework based on the programming language \texttt{C} and the 
parallelisation interface \texttt{MPI}. The Rayleigh quotient iteration is performed in double precision and each 
inversion is computed by a double precision flexible GMRES solver preconditioned with single precision algebraic multigrid. All 
results that we state in this section were obtained on the Juropa machine at J\"ulich Supercomputing Centre, a cluster 
with $2,\!208$ compute nodes, each with two Intel Xeon X5570 (Nehalem-EP) quad-core processors. This machine provides a 
maximum of $8,\!192$ cores for a single job. The code is compiled with the \texttt{icc}-compiler using the optimisation 
flags
\texttt{-O3}, \texttt{-ipo}, \texttt{-axSSE4.2} and \texttt{-m64}.

\begin{figure}[ptb]
\centering\scalebox{0.6}{\input{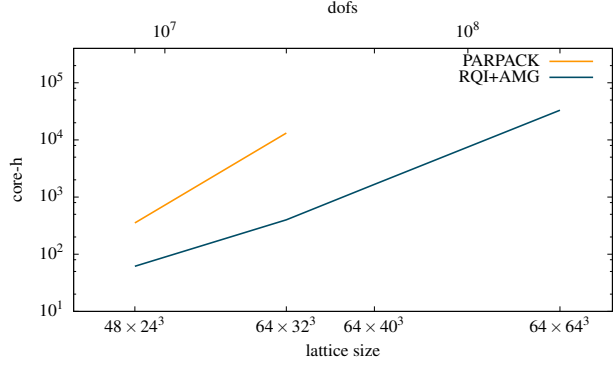}}
  \caption{Comparison of PARPACK and RQI+AMG, core hours needed to compute the $N_{eig}=20$ smallest eigenvectors of $Q$ for lattice sizes ranging from $48 \times 24^3$ to $64^4$ at constant physics.}
  \label{plot:parpack_vs_mgeigsolve}
\end{figure}

In Fig.~\ref{plot:parpack_vs_mgeigsolve} we compare the amount of core hours needed to compute the $N_{eig}=20$ smallest 
eigenvectors of $Q$ with RQI+AMG and with PARPACK~\cite{wwwPARPACK}. The latter is a publicly available parallel Arnoldi type 
eigensolver which is widely used in the lattice QCD community. It builds a Krylov subspace of a chosen dimension 
$N_{kv}$ and estimates $N_{eig}$ eigenvector approximations in this subspace. Thereafter the procedure is restarted, keeping 
the $N_{eig}$ approximations and improving them within a new subspace which consists of the $N_{eig}$ approximate eigenvectors 
and $N_{kv}-N_{eig}$ new vectors coming from a new Arnoldi iteration. In Fig.~\ref{plot:parpack_vs_mgeigsolve} we used 
$N_{kv} = 100$. We observe that RQI+AMG outperforms PARPACK by almost an order of magnitude already on rather small 
configurations of size $48 \times 24^3$. 
For a lattice volume of $64\times40^3$ PARPACK already exceeded the 24 hours job limit with 1024 
cores. 
The curves displayed in Fig.~\ref{plot:parpack_vs_mgeigsolve} also hints at that RQI+AMG 
scales better with the lattice size than PARPACK does. This is a major advantage for today's large volume simulations. 
Note that all configurations are at constant physics, i.e., they are all two flavour simulations at \(m_\pi \approx 
290\,\mathrm{MeV}\) and a lattice spacing \(a\approx0.071\,\mathrm{fm}\) (more details on the used configurations can 
be found in~\cite{Bali:2014gha}).

\begin{figure}[ptb]
\centering\scalebox{0.6}{\input{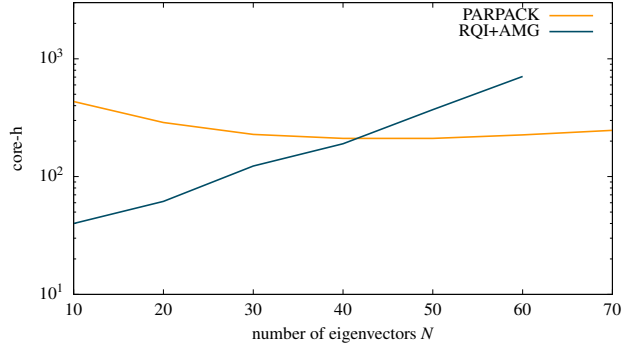}}
  \caption{Comparison of PARPACK and RQI+AMG, core hours as a function of the number of smallest eigenvectors $N_{eig}$ for a configuration size of $48 \times 24^3$.}
  \label{plot:parpack_vs_mgeigsolve_nmodes}
\end{figure}

However, the situation is less in favour of RQI+AMG when we investigate the scaling with the number of desired 
eigenvectors $N_{eig}$. In Fig.~\ref{plot:parpack_vs_mgeigsolve_nmodes} we compare the scaling with $N_{eig}$ for RQI+AMG and 
PARPACK. For PARPACK we invariably used $N_{kv}=200$ given that in our numerical experiments we did not see any significant difference in 
runtime when keeping $N_{kv}$ constant instead of taking $N_{kv}$ proportional to $N_{eig}$. Note that we always had at least $N_{eig} < 
\frac{1}{2}N_{kv}$. We observe that the runtime of RQI+AMG grows rapidly as the number of eigenvectors $N_{eig}$ is increased 
whereas PARPACK does not show any distinct dependence on $N_{eig}$.

Due to the orthonormalisation process in the Arnoldi procedure, PARPACK is expected to scale with 
$\mathcal{O}(N_{eig}^2)$. However for small numbers of eigenvectors $N_{eig}$, the number of restarts in PARPACK 
predominates the overall computations rather than the orthonormalisation process. Since we build all $N_{eig}$ current 
eigenvector approximations into the interpolation $P$ of algebraic multigrid,  
the corresponding coarse operator $Q_c = P^\dagger Q P$ has complexity $\mathcal{O}(N_{eig}^2)$, since each coarse lattice site 
holds $2N_{eig}$ variables which couple with each neighbouring coarse lattice site via a non-sparse $2N_{eig}\times2N_{eig}$ coupling matrix. Solving the coarse grid system 
is thus expected to scale at least as $\mathcal{O}(N_{eig}^2)$. In future work we plan to investigate an eigensolver approach 
wherein we do not need to build $P$ from all $N_{eig}$ current eigenvector approximations.

\begin{figure}[ptb]
\centering\scalebox{0.6}{\input{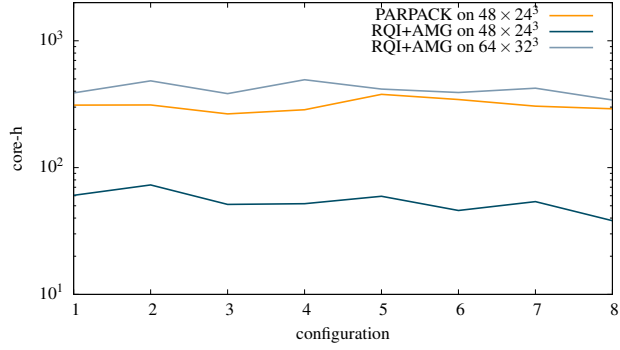}}
  \caption{Comparison of PARPACK and RQI+AMG, two ensembles at different lattice sizes, 8 statistically independent configurations each.}
  \label{plot:eig_hmc_traj1}
\end{figure}

Finally, in Fig.~\ref{plot:eig_hmc_traj1} we track the dependence of PARPACK and RQI+AMG on spectral fluctuations for eight statistically independent configurations for two different lattice volumes. We observe only minor fluctuations in core hours.

\section{Inexact Eigenmodes and Physics Application}
\label{sec:physicsapp}
We now use the above described eigensolver for low-mode averaging which is employed to improve the statistical signal of connected~\cite{DeGrand:2004qw,Giusti:2004yp} and disconnected~\cite{Neff:2001zr,Bali:2005fu,Foley:2005ac,Bali:2009hu} contributions to hadronic observables. In this work, we apply this method to pion- and eta-meson correlators.

Such noise reduction techniques are particularly important for quark line disconnected contributions. These arise, e.g., when calculating fermionic \(n\)-point functions of flavour-singlet quantities. For the eta-meson in the two-flavour (\(n_f=2\)) theory, for example, an interpolator is given by
\begin{equation} 
  O^\eta_x = \frac{1}{\sqrt{2}} \left( \bar{u}_x \gamma_5 u_x + \bar{d}_x\gamma_5 d_x\right)\text{.}
\end{equation}
Performing the Wick contractions to obtain the two-point function gives, in the case of mass-degenerate quarks,
\begin{align}
    C_{\eta}(x,y) = & \langle O^\eta_x \bar{O}^\eta_y \rangle \nonumber\\
  \propto & \tr\left(D^{-1}_{x,y} \gamma_5 D^{-1}_{y,x}\gamma_5 \right) - n_f \tr\left(D^{-1}_{x,x}\gamma_5\right)\tr\left(D^{-1}_{y,y}\gamma_5\right)\text{,}
  \label{eq:etacor}
\end{align}
where the first term is the connected part that can be calculated cheaply on a single source point \(y_0\) by using \(\gamma_5\)-hermiticity and translational invariance,
\begin{equation}
  \tr\left(D^{-1}_{x,y_0} \gamma_5 D^{-1}_{y_0,x}\gamma_5\right) = \tr\left(D^{-1}_{x,y_0}(D^{-1}_{x,y_0})^\dagger\right)\text{,}
\end{equation}
where the trace is over spin and colour indices.
For the disconnected contribution, however,
the propagator starts and ends at the same spacetime point and the calculation of the ``loop'' \(D_{x,x}^{-1}\gamma_5\) would require the inversion of the full matrix. In most cases this is computationally unrealistic due to the size of \(D\). Instead, one usually employs stochastic techniques for that (see~\cite{Bali:2009hu} and references therein), i.e., one calculates 
\begin{equation}
  Q^{-1} = D^{-1}\gamma_5 = \frac{1}{N_{stoch}} \sum_i^{N_{stoch}} \mid s_i \rangle \langle \eta_i \mid + \mathcal{O}(\sfrac{1}{\sqrt{N_{stoch}}})
  \label{eq:stochestim}
\end{equation}
for a sufficiently large number \(N_{stoch}\) of stochastic solutions \(\mid s_i \rangle\) of the linear system
\begin{equation}
  Q \mid s_i \rangle = \mid \eta_i \rangle\text{,}
\end{equation}
where \(\mid \eta_i \rangle\) is a random noise vector with the properties
\begin{equation}
  \frac{1}{N_{stoch}} \sum_i^{N_{stoch}} \mid \eta_i \rangle = \mathcal{O}(\sfrac{1}{\sqrt{N_{stoch}}})
  \quad\text{and}\quad
  \frac{1}{N_{stoch}} \sum_i^{N_{stoch}} \mid \eta_i \rangle \langle \eta_i \mid = \mathds{1} + \mathcal{O}(\sfrac{1}{\sqrt{N_{stoch}}})\text{.}
\end{equation}
A common choice, which we also use, is to draw the elements of \(\mid \eta_i \rangle\) from \(\mathbb{Z}_2 + i\mathbb{Z}_2\) noise.

As is clear from Eq.~\eqref{eq:stochestim}, this introduces additional stochastic noise for any finite number of estimates which adds to the gauge noise.
In other words, \(N_{stoch}\) must be chosen large enough so that the gauge noise dominates in the overall statistical error. 
This requires additional solves and becomes more expensive when going down to physical pion masses and large volumes, even with modern, e.g., multigrid based solvers.

To reduce the stochastic noise one can make use of various noise reduction techniques like partitioning~\cite{Bernardson:1993he,Viehoff:1997wi,Foley:2005ac}, the truncated solver method \cite{Bali:2009hu} or low-mode averaging (for this case known as truncated eigenmode acceleration)~\cite{Bali:2005fu,Foley:2005ac}, to name only a few. Which combination of these methods works best will in general not only depend on the efficiency of the solver but also on the observable under consideration. The eta-correlator is known to be low-mode dominated~\cite{Neff:2001zr}, therefore, it is the ideal quantity to test our new eigensolver and investigate the use of approximate eigenpairs in low-mode averaging.

\subsection{Low-Mode Averaging}
\label{sec:lma}
The basic idea of Low-Mode Averaging (LMA) is to split the operator, e.g., the Hermitian Dirac Operator \(Q = \gamma_5 D\), in two parts:
\begin{equation}
  Q^{-1} = Q^{-1}_{low} + Q^{-1}_{high}\text{,}
\end{equation}
where \(Q_{low}^{-1}\) contains the the contributions to \(Q^{-1}\) from the \(N_{eig}\) lowest eigenmodes:
\begin{equation}
    Q^{-1}_{low} = \sum_{i}^{N_{eig}} \frac{1}{\lambda_i} \mid v_i \rangle\langle v_i \mid\text{.}
  \label{eq:qlowinv}
\end{equation}
\(Q^{-1}_{high}\) is the remaining part of \(Q^{-1}\).

For the eta-correlator (cf. Eq.~\eqref{eq:etacor}), low-mode averaging works as follows: We need to calculate both the connected (pion) correlator \(C_{con}(x,y) = \tr (Q_{x,y}^{-1}Q_{y,x}^{-1})\) and the disconnected contribution \(C_{dis}(x,y) = \tr(Q_{x,x}^{-1}) \tr(Q_{y,y}^{-1})\). LMA can be used for both terms:
The connected term, averaged over the spatial volume and only depending on the Euclidean time distance \(t\), reads
\begin{equation}
  C_{con}(t) = C_{con}^{low}(t) + C_{con}^{high}(t) = C_{con}^{low}(t) + \left(C_{con}^{p2a}(t) - C_{con}^{low,p2a}(t)\right)\text{,}
  \label{eq:conlma}
\end{equation}
where with \(x = (\vect{x}, t_0 + t)\text{, } y = (\vect{y}, t_0)\text{, } y_0 = (\vect{y}_0, t_0)\) the individual terms are given as
\begin{align}
  \label{eq:corconlow}
  C_{con}^{low}(t) = & \frac{1}{V} \sum_{\vect{x},\vect{y},t_0} \tr \left[ (Q_{low}^{-1})_{x,y}(Q_{low}^{-1})_{y,x}\right]\text{,}\\
  \label{eq:corconlowpt2a}
  C_{con}^{low,p2a}(t) = & \sum_{\vect{x}} \tr \left[ (Q_{low}^{-1})_{x,y_0}(Q_{low}^{-1})_{y_0,x}\right]\text{,}\\
  \label{eq:corconpt2a}
C_{con}^{p2a}(t) = & \sum_{\vect{x}}\tr\left[\left(D^{-1}\right)_{x,y_0} \left(D^{-1}\right)^\dagger_{x,y_0}\right]\text{.}
\end{align}
The splitting is performed in a straight-forward way: First, we calculate the low-mode contribution \(C_{con}^{low}\) which uses the full (all-to-all) information contained in the eigenmodes. The high-mode correction (the terms in the brackets of Eq.~\eqref{eq:conlma}) is calculated from the exact point-to-all twopoint function \(C_{con}^{p2a}\) and \(C_{con}^{low,p2a}\) obtained from the eigenmodes at point \(y_0\).

For the disconnected terms, we correlate two loops at times \(t_0\) and \(t_0+t\),
\begin{equation}
  C_{dis}(t) = \frac{1}{N_t} \sum_{t_0} L(t_0+t) L(t_0)\text{,}
  \label{eq:dislma}
\end{equation}
where low-mode substitution is applied to the calculation of the individual loops:
\begin{equation}
  L(t) = \sum_{\vect{x}}\tr\left[ Q^{-1}_{x,x} \right] = L^{low}(t) + L^{high}(t)\text{.}
\end{equation}
Again, we calculate the low-mode contribution using Eq.~\eqref{eq:qlowinv},
\begin{equation}
  L^{low}(t) = \sum_{\vect{x}} \tr \left[ (Q_{low}^{-1})_{x,x} \right]\text{.}
  \label{eq:loophigh}
\end{equation}
To remove the high modes from \(Q\), we use the orthonormal projector 
\begin{equation}
  \proj{P} = \mathds{1} - \sum_{i}^{N_{eig}} \mid v_i \rangle \langle v_i \mid\text{,}
  \label{eq:loophighprojector}
\end{equation}
and estimate
\begin{equation}
  L^{high}(t) = \sum_{\vect{x}} \tr \left[ \left( \proj{P} Q \right)^{-1}_{x,x} \right]
  \label{eq:dislmahigh}
\end{equation}
stochastically. Note that \(\proj{P}Q = Q\proj{P}\) in this case.

If the low-modes dominate, as is the case for the eta-correlator, less estimates in the calculation of \(Q^{-1}_{high}\) are required to achieve a given accuracy.
However, the overall cost can be dominated by the calculation of the eigenmodes.
Therefore, low-mode averaging is often only cost-effective if the eigenmodes can be reused many times.
\subsection{Approximate Low-Mode Averaging}
\label{sec:ainv}
Besides the development of faster eigensolvers, it is also possible to reduce the cost of LMA by relaxing the eigenmode tolerance in the eigensolver and then correct for this reduced accuracy.

We denote the \(i\)-th approximate eigenpair of \(Q\) as \((\tilde{\lambda}_i, \mid \tilde{v}_i \rangle)\) with the eigenmode accuracy 
\begin{equation}
  \epsilon_i = \norm{Q \mid \tilde{v}_i \rangle - \tilde{\lambda}_i \mid \tilde{v}_i \rangle}\text{,}
\end{equation}
and assume that the \(\mid \tilde{v}_i\rangle\) are orthonormalised.
Further, we define a matrix
\begin{equation}
    A_{ij} = \langle \tilde{v}_i \mid Q \mid \tilde{v}_j \rangle\text{,}
\end{equation}
which we will use to account for the inexactness of the eigenmodes:
With \(\mid \delta v_i \rangle \) denoting the error of the approximate eigenvector we have
\begin{equation}
  \mid \tilde{v}_i \rangle = \mid v_i \rangle + \mid \delta v_i \rangle\text{.}
  \label{eq:inexactevec}
\end{equation}
Writing \(A\) in these terms
\begin{equation}
    A_{ij} = \lambda_j \delta_{ij} + \lambda_i \langle v_i \mid \delta v_j \rangle + \lambda_j \langle \delta v_i \mid v_j\rangle + \langle \delta v_i \mid Q \mid \delta v_j \rangle\text{,}
    \label{eq:longa}
\end{equation}
shows that \(A\) would be diagonal if the eigenmodes were exact, i.e., if all \(\delta v_i = 0\). Using the inverse of \(A\) instead of the inverse eigenvalues in Eq.~\eqref{eq:qlowinv} the expression for the low-mode part of the Hermitian propagator generalises to
\begin{equation}
    \tilde{Q}_{low}^{-1} = \sum_{i,j}^{N_{eig}} \left(A^{-1}\right)_{ij} \mid \tilde{v}_i \rangle \langle \tilde{v}_j \mid\text{.}
  \label{eq:qlowinvapprox}
\end{equation}
Using this corrected inverse for LMA amounts to replacing \(Q_{low}^{-1}\) by \(\tilde{Q}_{low}^{-1}\) in Eqs.~\eqref{eq:corconlow} to \eqref{eq:corconpt2a} and Eq.~\eqref{eq:loophigh}. Note that in the exact case Eq.~\eqref{eq:qlowinvapprox} is identical to Eq.~\eqref{eq:qlowinv}. 
The high mode part is now obtained by replacing $\proj{P}$ in Eq.~\eqref{eq:qlowinv} by an oblique projection
\begin{equation}
  \tilde{\proj{R}} = \mathds{1} - \sum_{i,j}^{N_{eig}} Q \mid \tilde{v}_i \rangle (A^{-1})_{ij} \langle \tilde{v}_j \mid\text{.}
    \label{eq:approxproj}
\end{equation}

Taking \(A^{-1}\) instead of the inverse eigenvalues improves the estimate for \(Q_{low}^{-1}\) by combining information from the full space spanned by the inexact eigenvectors.
Note that \(\tr\left(Q \tilde{Q}_{low}^{-1}\right) = 12\,N_{eig}\). When combining Eq.~\eqref{eq:qlowinvapprox} and Eq.~\eqref{eq:approxproj} together either with Eq.~\eqref{eq:dislmahigh} or Eq.~\eqref{eq:conlma} we still obtain unbiased results.
\subsection{Use of Multigrid Test Vectors in LMA}
\label{sec:cutoff}
One can go even one step further: Studying the multigrid approach, we observe that the test vectors can already be quite accurate
eigenvectors directly after our initial multigrid setup phase. In the ensemble considered here, the precision after 30 setup iterations is of the order \(\epsilon \approx 10^{-3}\) and it turns out that we can indeed use the test vectors as approximate eigenmodes for LMA.

At the beginning the multigrid test vectors are initialised to random vectors and, 
since the ensemble average of \(\langle r \mid Q \mid r \rangle\) vanishes (where \(\mid r \rangle\) is a random vector), this has the consequence, that some of the initial eigenvalues are systematically underestimated and the resulting approximate eigenmodes do not respect the mass gap. Including these will obviously affect the quality of \(Q^{-1}_{low}\).
\begin{figure}[ptb]
  \centering
  \resizebox{0.8\textwidth}{!}{
  \includegraphics{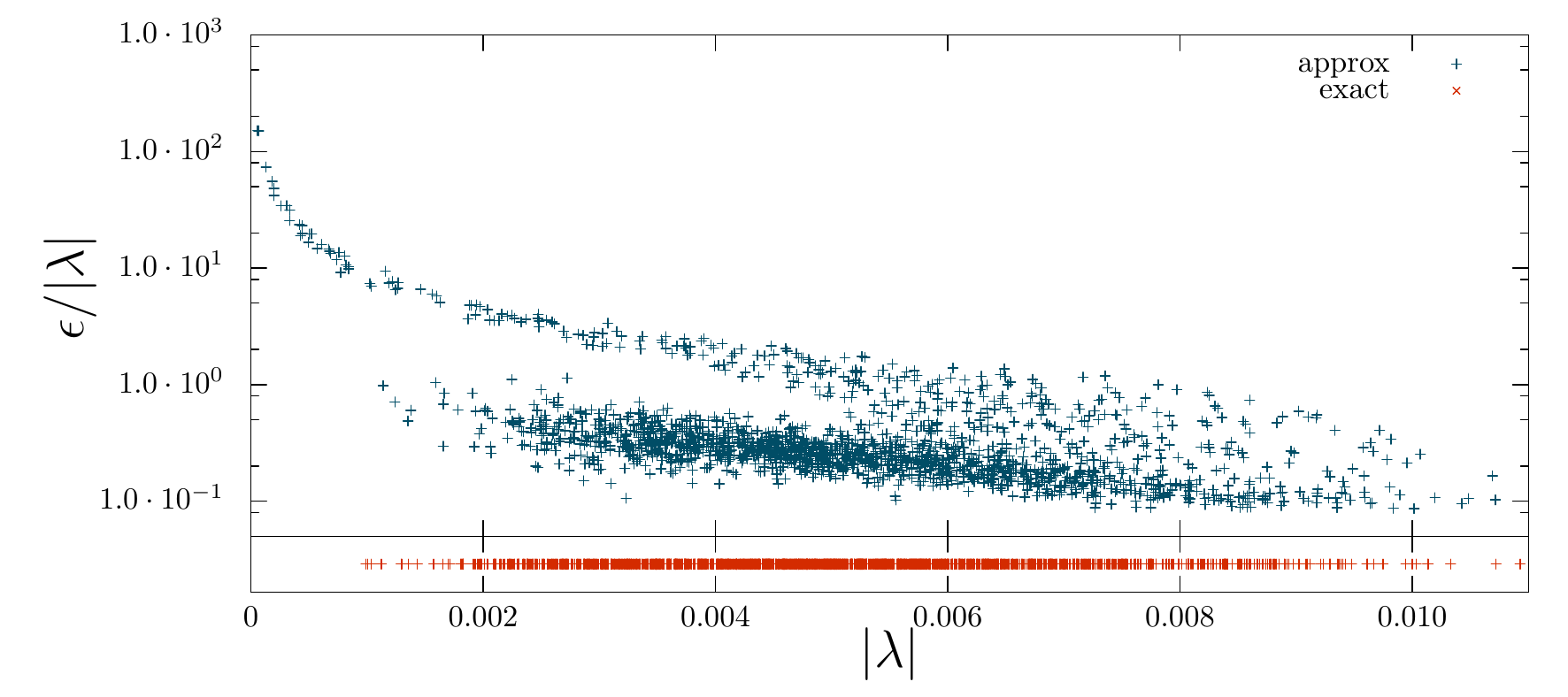}
  }
  \caption{The lowest 30 eigenvalues of \(Q\) on each of 64 configurations on a \(40^3\times 64\) volume at \(m_\pi \approx 290\,\textrm{MeV}\). The bottom section shows the exact (\(\epsilon \leq 10^{-8}\)) spectrum. The upper plot relates the eigenvalues taken directly after the Multigrid setup with their relative accuracies \(\frac{\epsilon}{|\lambda|}\).}
  \label{fig:eigenvaluespectrum}
\end{figure}

Luckily, such deviations can be detected by studying the relative precision \(\sfrac{\epsilon}{|\tilde{\lambda}|}\) of the eigenmodes, see Fig.~\ref{fig:eigenvaluespectrum}. It turns out that the eigenvalues not respecting the mass gap are the ones with large \(\sfrac{\epsilon}{|\tilde{\lambda}|}\). By imposing a cutoff, we restrict the set of eigenpairs used in our computation,
\begin{equation}
  \left\{(\tilde{\lambda}, \mid \tilde{v} \rangle, \epsilon)\phantom{\bigg.}\right\} \to \left\{(\tilde{\lambda}, \mid \tilde{v} \rangle, \epsilon ) \quad \bigg| \quad\frac{\epsilon}{|\tilde{\lambda}|} \leq c\right\}\text{.}
\end{equation}
Any normalisation cancels from the above ratio making this choice of cutoff independent of lattice volume and pion mass. Once a reasonable cutoff value \(c\) has been found, it can be used on different ensembles.

\begin{figure}[ptb]
  \centering
  \resizebox{0.8\textwidth}{!}{
  \includegraphics{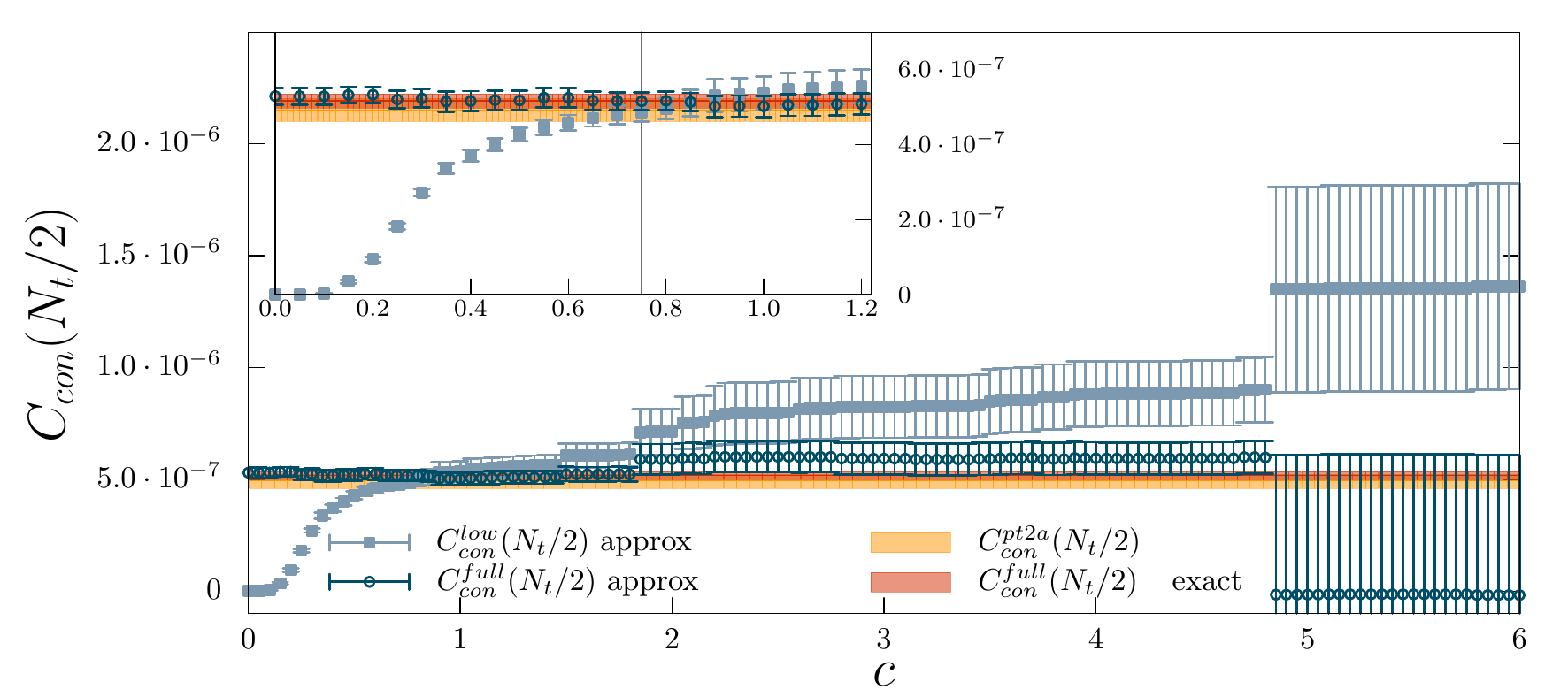}
  }
  \caption{The connected pseudoscalar two-point correlator at \(t = \sfrac{N_t\,a}{2}\) using only the eigenmodes that have a relative error smaller than \(c\). The boxes show the contribution coming only from the inexact low-modes, whereas the circles show the corrected two-point function. For comparison, the red bar shows the value that can be obtained when using 20 exact eigenmodes with \(\epsilon \leq 10^{-8}\) and the yellow bar marks the conventional point-to-all result.}
  \label{fig:varyCutoff}
\end{figure}

An ideal benchmark quantity for selecting the cutoff is the low-mode contribution to the connected two-point function: It can be calculated without stochastic estimation, i.e., it has gauge fluctuations only and the calculation is cheap. Fig.~\ref{fig:varyCutoff} shows how \(C_{con}^{low}\) varies with the cutoff. The depicted data are the values at the central timeslice where the relative contributions of the low-modes are largest and therefore the effect of using inexact modes can be detected most easily.
%
If \(c\) is chosen too big we encounter large errors both in \(C_{con}^{low}\) and \(C_{con}^{full}\) due to the weight given to some irrelevant directions by underestimated eigenvalues.
In contrast, if \(c\) is small, we will not benefit from LMA.
In any case, after adding the high-mode correction (cf.~Eq.~\eqref{eq:conlma}), within errors we always obtain the correct result, as shown by the horizontal bands for the point-to-all and the LMA case.
We find that a cutoff of \(c=0.75\) is a good compromise. 

Again we stress that, although the low-mode part is affected by both the number and the accuracy of the eigenmodes, the full, high-mode corrected quantity is stable and unbiased. Varying the cutoff empirically demonstrates the validity of this statement.
\subsection{Results}
\label{sec:results}
For a first practical test, we use the same ensemble as in the previous sections: A moderately large volume of \(V = 40^3\times 64\) with two sea quark flavours generated by QCDSF at a pion mass of \(m_\pi \approx 290\,\mathrm{MeV}\) (\(L\,m_\pi \approx 4.19\)) and an inverse coupling \(\beta = 5.29\) (corresponding to a lattice spacing \(a\approx 0.071\,\mathrm{fm}\)), see, e.g., \cite{Bali:2014gha} for the simulation details. This allows us to explore the methods at moderate costs but under real-world conditions.
To reduce excited states contributions we use 400 steps of Wuppertal smearing~\cite{Gusken:1989ad} with smearing parameter \(\delta = 0.25\) for the quark sources and sinks and also for the eigenvectors. The gauge field used within the quark smearing is APE-smeared \cite{Falcioni:1984ei} with weight factor \(\alpha = 0.25\).

We perform our calculations on 64 independent configurations. On each of these we compute 30 inexact eigenmodes, just performing the multigrid setup with 30 steps and no further Rayleigh quotient iteration on the test vectors. These inexact eigenmodes have an accuracy of typically \(\epsilon \approx 10^{-3}\). On average, three out of the 30 modes are excluded by our cutoff choice \(c=0.75\). To compare and verify the inexact results, we also compute the smallest 20 eigenmodes using the algorithm described in Sec.~\ref{sec:algo} with a tolerance of \(\epsilon = 10^{-8}\). We refer to them as ``exact''.

\begin{figure}[p]
  \resizebox{0.5\textwidth}{!}{
  \includegraphics{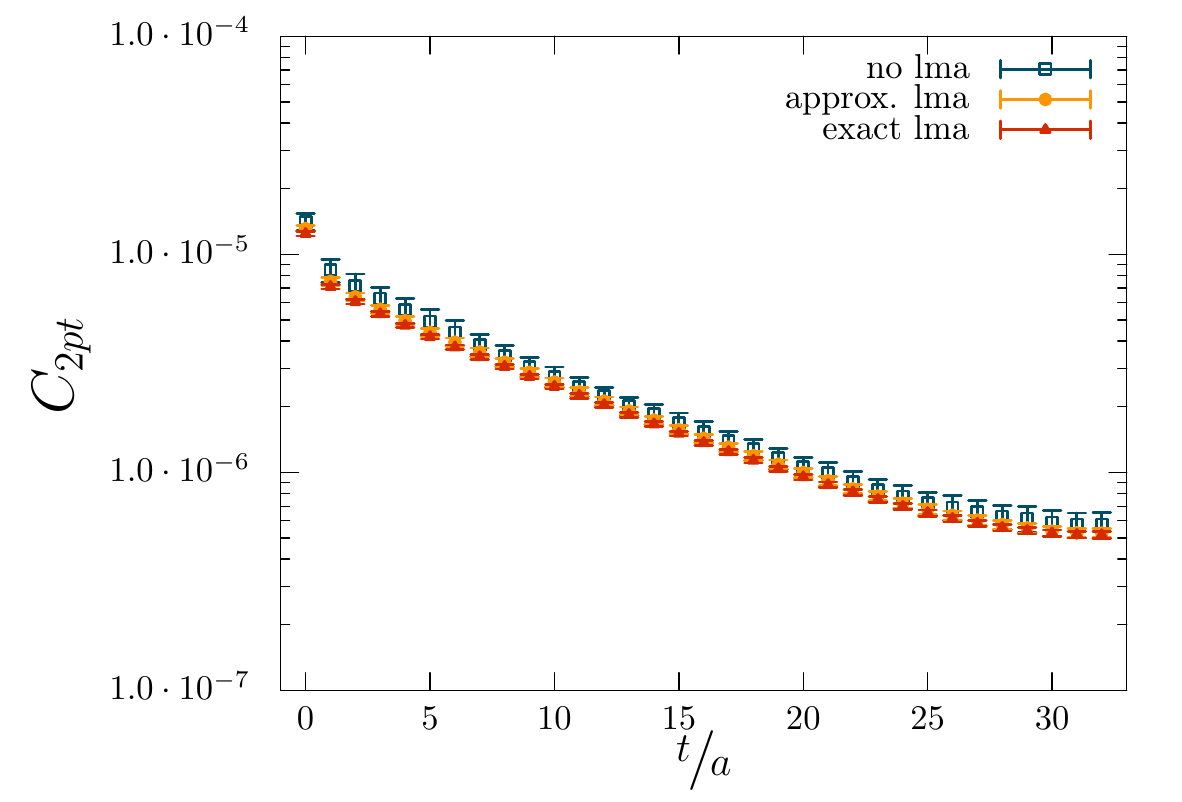}
  }
  \resizebox{0.5\textwidth}{!}{
  \includegraphics{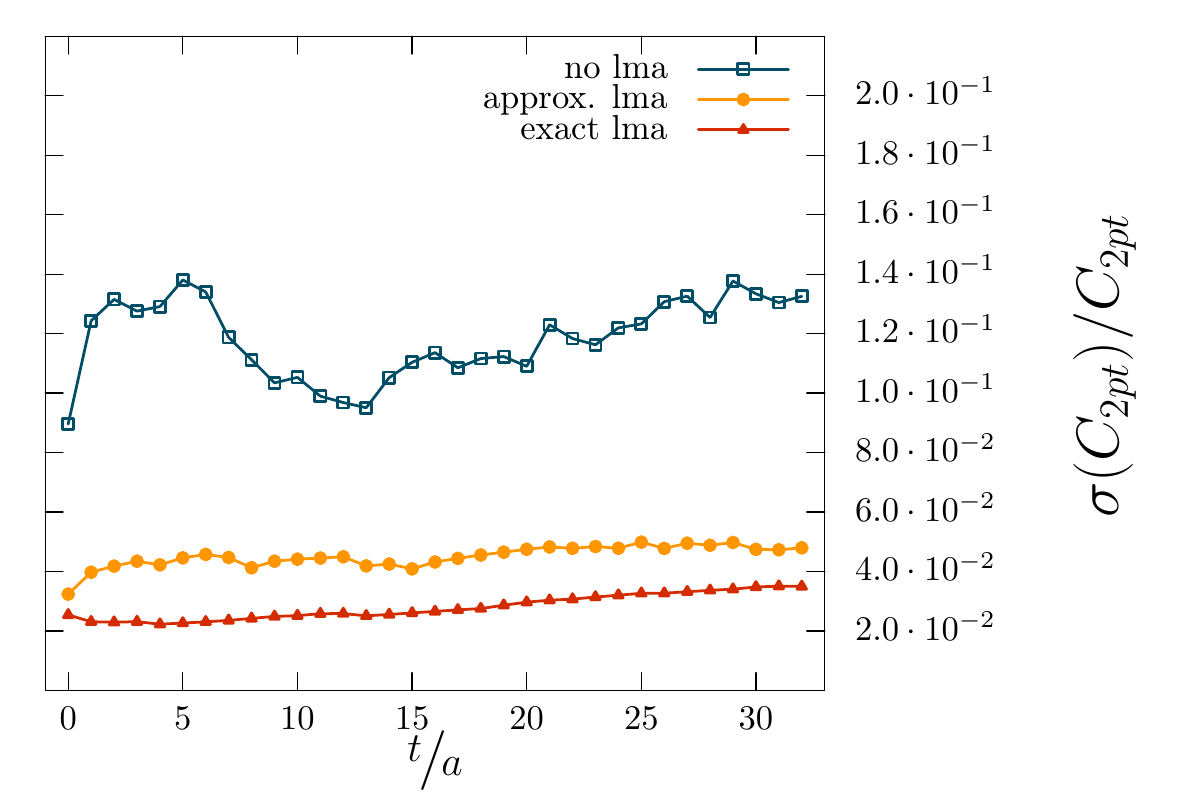}
  }
  \caption{The connected pseudoscalar (pion) correlator (left) and its relative error at each timeslice (right), calculated with exact (red triangles), inexact (orange circles) and without (blue boxes) LMA.}
  \label{fig:conncorrelators}
\end{figure}

\begin{figure}[p]
  \resizebox{0.5\textwidth}{!}{
  \includegraphics{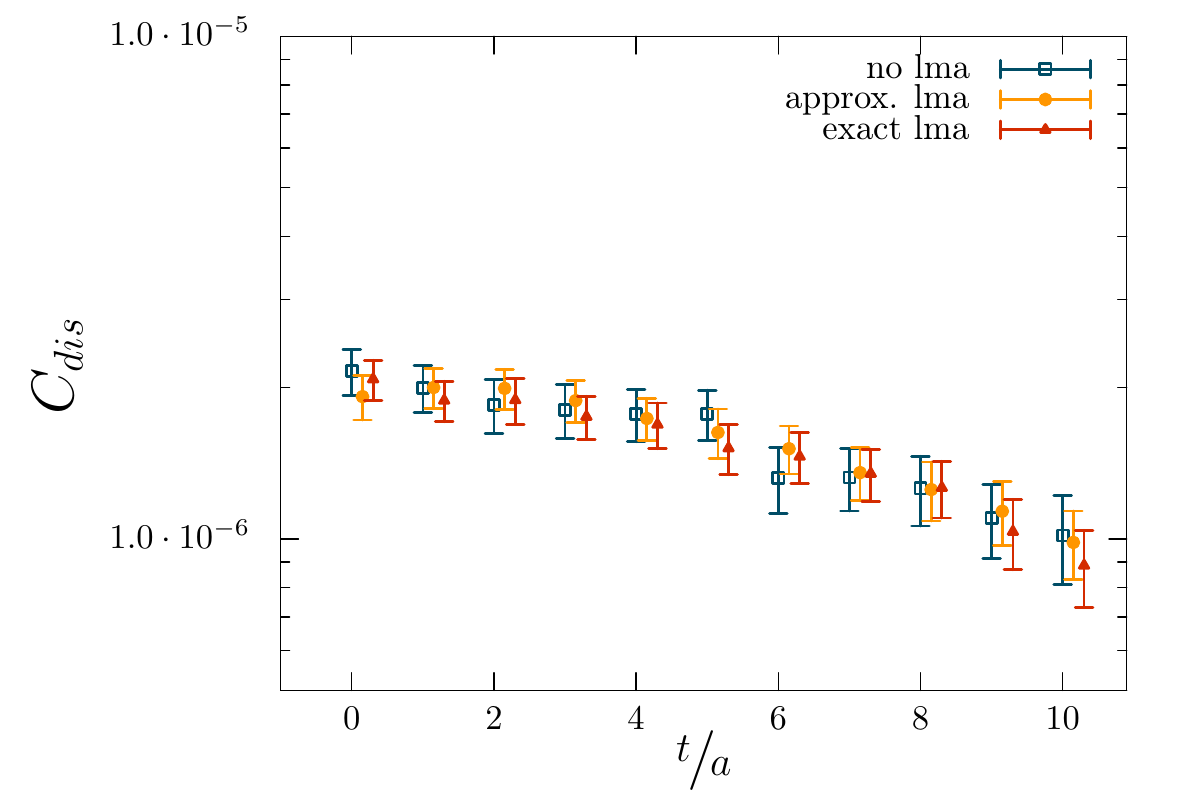}
  }
  \resizebox{0.5\textwidth}{!}{
  \includegraphics{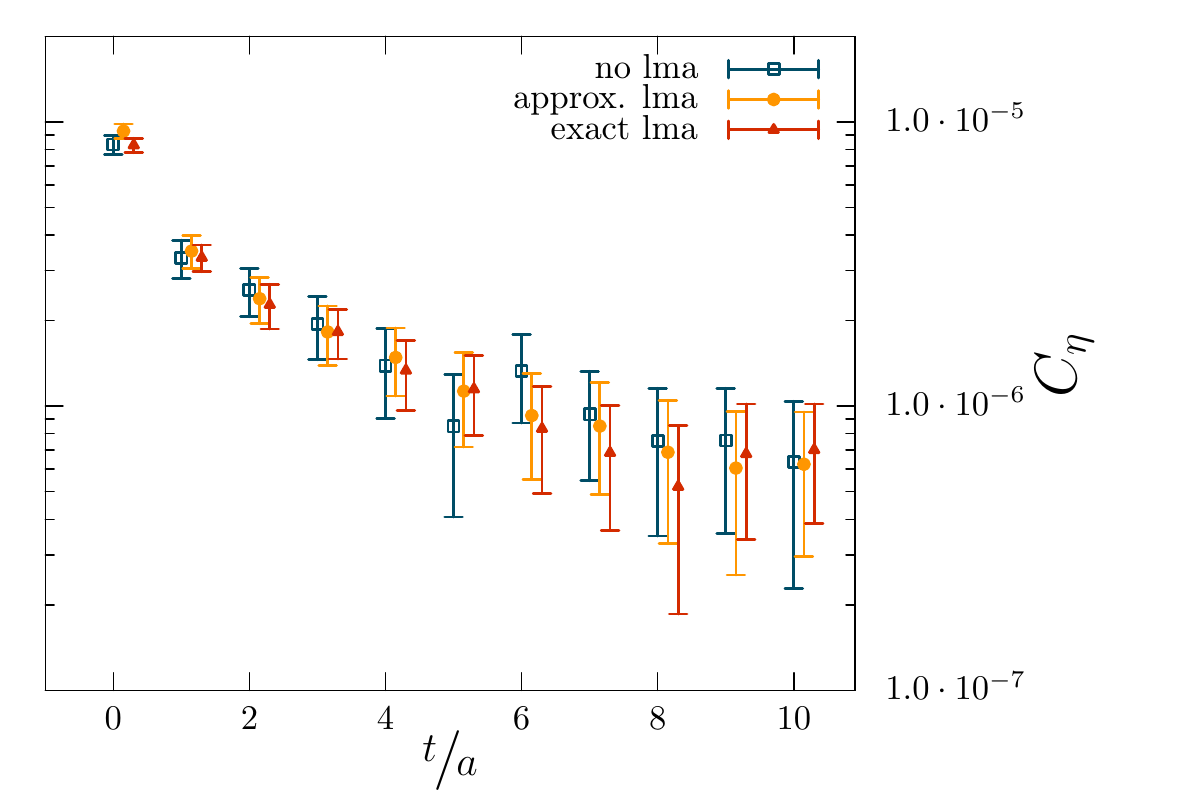}
  }
  \caption{The disconnected contribution (left) and the full eta-correlator (right), calculated with exact (red triangles), inexact (orange circles) and without (blue boxes) LMA, each using \(N_{stoch}=20\) stochastic estimates. The points are slightly shifted horizontally to improve visibility.}
  \label{fig:discorrelators}
\end{figure}

Fig.~\ref{fig:conncorrelators} shows the connected (pion) correlator and its relative errors. Due to the spatial volume averaging, LMA works very well in this case. The connected contribution gives already a first indication that our improvement techniques work: We obtain nearly the same errors with approximate LMA as with exact eigenmodes.
For the disconnected contribution, we employ time dilution~\cite{Viehoff:1997wi,O'Cais:2004ww,Bali:2014pva} with a distance \(\Delta t = 4\,a\) in all cases. 
In this case the data of both exact and approximate LMA agree with the points for which no low-mode averaging has been used as can be seen from Fig.~\ref{fig:discorrelators}. The central values show a slightly smoother behaviour. When combining the connected and the disconnected data to form the eta-correlator in Fig.~\ref{fig:discorrelators}, the errors increase towards larger times and we find that the data do not obey the expected exponential decay. This is probably an effect of our limited statistics and insufficient sampling of topological sectors~\cite{Bali:2014pva}.

\begin{figure}[p]
  \resizebox{0.5\textwidth}{!}{
  \includegraphics{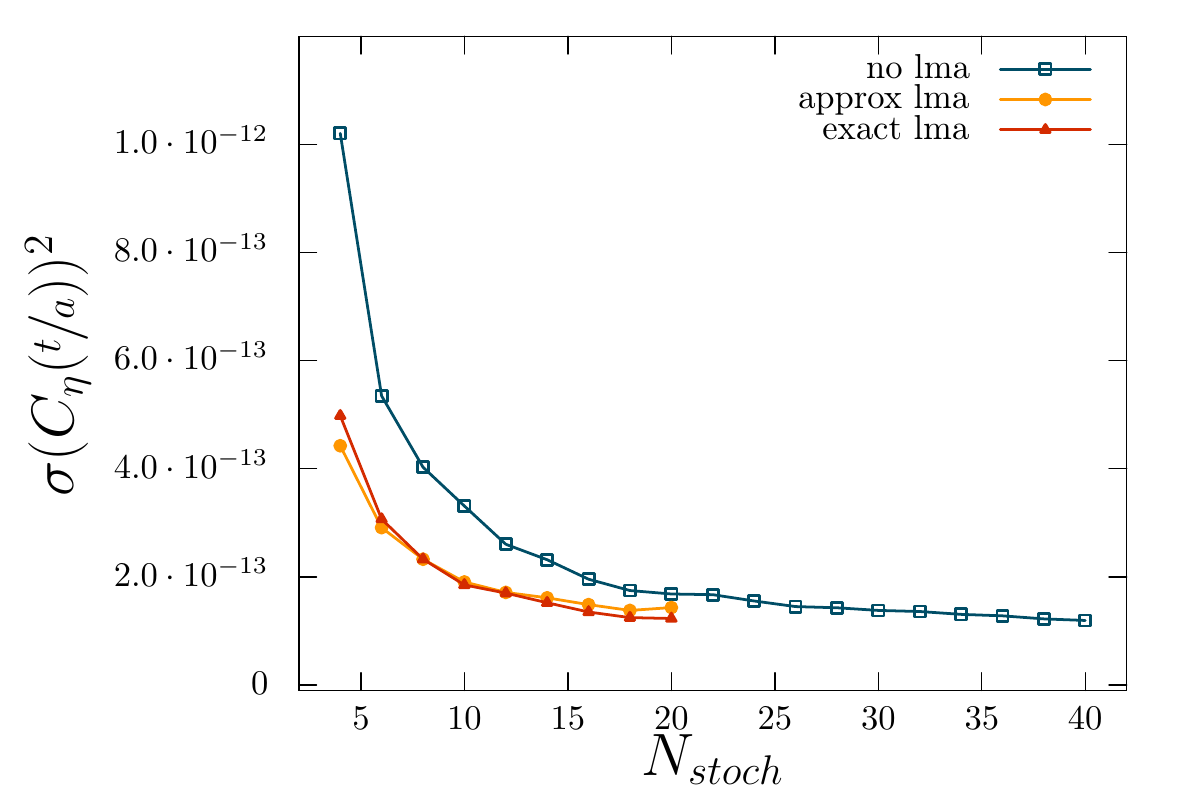}
  }
  \resizebox{0.5\textwidth}{!}{
  \includegraphics{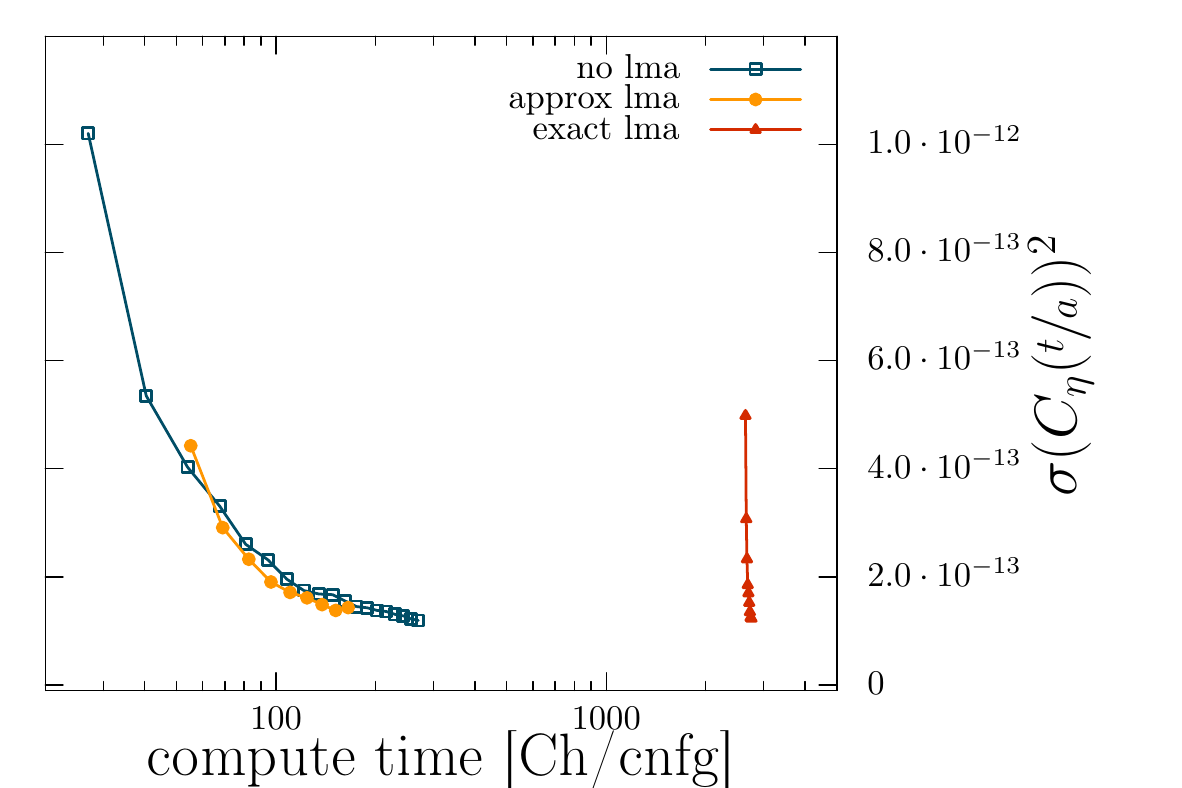}
  }
  \caption{The average quadratic error of the first ten timeslices of the eta-correlator using exact (red triangles), inexact (orange circles) and no low-mode averaging (blue boxes). The left plot shows how many stochastic estimates \(N_{stoch}\) are needed to obtain a certain error (on our limited statistics of 64 configurations). The right plot compares the actual cost on SuperMUC at LRZ, taking the time for the calculation of the eigenmodes and the stochastic estimation into account.}
  \label{fig:costploteta}
\end{figure}

The previous plots show that both exact and inexact LMA reduce the errors. Alternatively, one could of course also increase the number of stochastic estimates \(N_{stoch}\) so that LMA is not necessary. Fig.~\ref{fig:costploteta} shows the quadratic error, averaged over the first ten time slices -- after that point the noise grows rapidly -- depending on the number of stochastic solves.
This comparison shows the efficiency of LMA: In all cases, roughly twice as many inversions are necessary without LMA. Approximate and exact LMA are nearly equivalent. Finally, the most interesting question is how the methods compare at fixed computational cost. Fig.~\ref{fig:costploteta} shows the actual compute time needed for one configuration to obtain a certain error in the eta-correlator. It turns out that using the described methods we can reduce the cost of LMA by a factor of roughly ten. This speed-up means approximate LMA is cost-efficient and feasible for the use in large-scale computations, even if only a small number of different \(n\)-point functions needs to be computed.

\section{Conclusions and Outlook}
\label{sec:conclusions}
We made DD-$\alpha$AMG available for $Q$ by replacing the domain decomposition smoother by GMRES and integrated it into a Rayleigh quotient iteration in order to compute the smallest eigenpairs of $Q$. For moderate numbers of eigenpairs we obtained speed-ups of roughly an order of magnitude over PARPACK. We plan to work on improving the scaling with the number of eigenpairs and furthermore to explore possible benefits by incorporating multigrid in other shift-and-invert based eigensolvers.

Moreover, when just using the multigrid setup test vectors, our improved techniques for inaccurate eigenpairs enable us to get a decent signal for the pion- and eta-correlator that is comparable to the result one obtains when using exact eigenpairs, at only a fraction of the cost.
Of course whether this is possible or whether one needs a few iterations on the eigenvector approximations will depend on the observable and also strongly on the volume and pion mass. 

For smaller pion masses when inversions become more expensive and the spectrum is even more low-mode dominated the speed-up may be even more pronounced. We will report on this in a forthcoming publication. Also, the required number of eigenmodes at different volumes and an optimal tuning of the multigrid setup will be investigated, steps that are needed to get an optimal speed-up from which we will benefit in future high-statistics measurements.
%
%

{\small
\bigskip
This work is partially funded by Deutsche Forschungsgemeinschaft (DFG) within the transregional collaborative research centre 55 (SFB-TRR55). All numerical results in Sec.~\ref{sec:scaling} were obtained on Juropa at J\"ulich Supercomputing Centre (JSC) through NIC grant HWU12.
We also acknowledge computer time on SuperMUC at the Leibniz Rechenzentrum in Garching.
}
\bibliographystyle{apsrev}
{\small\bibliography{bibliography}}
\end{document}